\documentclass[12pt]{article}

\usepackage[a4paper,left=30mm,right=20mm,top=20mm,bottom=20mm]{geometry}
\usepackage{graphics}
\usepackage{amsmath}
\usepackage{epsfig}
\usepackage{cite}

\title{Phenomenology of $k_T$-factorization for inclusive \\Higgs boson production at LHC}

\author{A.V.~Lipatov$^{1,\,2}$, M.A.~Malyshev$^1$, N.P.~Zotov$^1$}

\begin{document}

\maketitle

\begin{center}

{\it $^1$Skobeltsyn Institute of Nuclear Physics, Lomonosov Moscow State University, 119991 Moscow, Russia}\\
{\it $^2$Joint Institute for Nuclear Research, Dubna 141980, Moscow Region, Russia}

\end{center}

\vspace{0.5cm}

\begin{center}

{\bf Abstract }

\end{center} 

\indent
We investigate the inclusive Higgs boson production in proton-proton collisions 
at high energies in the framework of $k_T$-factorization QCD approach. 
The attention is focused on the dominant off-shell gluon-gluon fusion subprocess 
$g^* g^* \to H \to \gamma \gamma$, where the transverse momentum of 
incoming gluons are taken into account. The transverse momentum dependent 
(or unintegrated) gluon densities of the proton are determined using the CCFM evolution equation 
as well as the Kimber-Martin-Ryskin prescription.  
We study the theoretical uncertainties of our
calculations and perform the comparison with the results of 
traditional pQCD evaluations.
Our predictions agree well with the first experimental data taken 
by the ATLAS collaboration at the LHC.
We argue that further studies of the Higgs boson production are
capable of constraining the
unintegrated gluon densities of the proton.

\vspace{1.0cm}

\noindent
PACS number(s): 12.38.Bx, 14.80.Bn

\newpage
\indent

In 2012, during the search for the SM Higgs boson at the LHC,
the CMS and ATLAS collaborations observed a new particle\cite{1,2},
undiscovered before. Some time later, with additional data and improved analysis strategy,
a specific spin-2 hypothesis has been excluded with a confidence
level above 99.9\%\cite{3}. Spin and relative production rates of the observed new particle in  
different decay modes conform to the SM expectations for the Higgs boson.
So, the experimental detection of the Higgs particle has been claimed
giving us the confidence in the physical picture of fundamental interactions 
which follows from the SM Lagrangian. Endeed, it has become
a great triumph of the SM and marked new stage in high energy physics.

Very recently the ATLAS collaboration has reported first measurements
of the Higgs boson differential cross sections in the diphoton decay mode\cite{4}.
In particular, the distributions on the diphoton transverse momentum $p_T^{\gamma \gamma}$, rapidity
$|y^{\gamma \gamma}|$ and helicity angle $|\cos \theta^*|$ have been 
presented\footnote{The helicity angle $\theta^*$ is defined as the angle between the beam axis
and the photons in the Collins-Soper frame of the Higgs boson.}.
These observables describe the fundamental kinematic properties 
of the discovered Higgs particle, probe its spin and test the 
corresponding theoretical calculations within the QCD,
which are performed 
for basic gluon-gluon fusion subprocess $gg \to H$\cite{5,6,7,8,9,10}
at next-to-next-to-leading order (NNLO)\cite{10,11,12,13,14,15}
and matched with soft-gluon resummation carried out up 
to next-to-next-to-leading logarithmic 
accuracy (NNLL)\cite{16,17}.
The measured cross sections are higher than the central SM expectations,
although no significant deviations from the SM predictions are observed
within the experimental and theoretical uncertainties\cite{4}.

In the present note we analyse recent ATLAS data using
the $k_T$-factorization QCD approach\cite{18,19}.
This approach is based on the 
Balitsky-Fadin-Kuraev-Lipatov (BFKL)\cite{20} or Ciafaloni-Catani-Fiorani-Marchesini 
(CCFM)\cite{21} gluon evolution equations and provides solid theoretical grounds for the 
effects of initial gluon radiation and intrinsic gluon transverse momentum. 
A detailed description and discussion of the $k_T$-factorization
formalism can be found, for example, in reviews\cite{22}.
Here we only mention that the soft gluon resummation formulas implemented in 
NNLL calculations\cite{16,17} 
are the result of the approximate treatment of the solutions
of CCFM equation\cite{23}.
Previously, the $k_T$-factorization formalism 
has been applied\cite{23,24,25,26} to calculate transverse momentum distribution of the 
inclusive Higgs boson production. The effective Lagrangian\cite{27,28}
for the Higgs boson coupling to gluons (valid in the large top quark mass limit) 
has been used to calculate the amplitude of basic gluon-gluon fusion subprocess.
However, the calculations\cite{23,24} neglect the transverse momentum of
initial gluons in the production amplitude, and the simplified
solution of the CCFM equation in the single loop approximation 
(where the small-$x$ effects are omited) has been applied in\cite{23}.
In\cite{24} the transverse momenta of incoming gluons are generated 
at the last evolution step (the so-called Kimber-Martin-Ryskin prescription).
The off-shell gluon-gluon fusion amplitude $g^*g^* \to H$
has been evaluated independently in\cite{25,29}, and corresponding 
expression\cite{29} has been implemented\cite{26} in a Monte Carlo event generator \textsc{cascade}\cite{30}.
The investigations\cite{23,24,25,26,29} inspired further studies\cite{31,32,33} of the inclusive 
Higgs boson production where finite mass effects in the triange quark loop 
have been taken into account.
The associated production of the Higgs boson and beauty or top quark pair at high energies 
in the $k_T$-factorization approach have been also investigated in\cite{34}.

Our present consideration is based mainly on the study\cite{25}
where the inclusive and jet associated Higgs boson cross sections
have been investigated. We calculate the off-shell amplitude of gluon-gluon fusion
subprocess $g^*g^*\to H\to \gamma \gamma$, which is not provided by the 
previous papers\cite{23,24,25,26,29,31,32,33}, and evaluate the total and differential sections of the 
Higgs boson production (in the diphoton decay mode). Numerically, we apply the unintgrated,
or transverse momentum dependent (TMD) gluon densities of the proton obtained 
from the numerical solution of CCFM evolution equation\cite{35,36}. 
We analyze the transverse momentum $p_T^{\gamma \gamma}$, 
rapidity $|y^{\gamma \gamma}|$ and helicity angle $|\cos \theta^*|$
distributions measured by the ATLAS collaboration and compare our 
predictions with the data\cite{4}. 
Such calculations in the framework of $k_T$-factorization QCD approach 
are performed for the first time.
Additional motivation of our study is that the transverse momentum of 
Higgs boson is strongly related to the initial gluon transverse momenta\cite{25},
and therefore such an observable could impose constraint on the TMD gluon density of the  
proton\footnote{Recently a program of QCD measurements at the LHC 
has been proposed where the Higgs boson is considered as a gluon trigger\cite{37}.}. 

Let us start from a short review of calculation steps. We describe 
first the evaluation of the off-shell $g^*g^* \to H \to \gamma \gamma$
production amplitude. In the limit of large top quark mass 
$m_t \to \infty$ the effective Lagrangian for the Higgs boson 
coupling to gluons reads\cite{27,28}
\begin{equation}
  {\cal L}_{ggH} = {\alpha_s\over 12\pi} \left(G_F \sqrt 2\right)^{1/2} G_{\mu \nu}^a G^{a\, \mu \nu} H,
\end{equation}

\noindent 
where $G_F$ is the Fermi coupling constant, $G_{\mu \nu}^a$ is the gluon field strength tensor and
$H$ is the Higgs scalar field. The large $m_t$ approximation is valid 
to an accuracy of $\sim 5$\% in the mass range $m_H < 2 m_t$, and,
of course, is applicable at the measured Higgs boson mass $m_H \sim 125$~GeV\cite{1,2,3,4}.
The triangle vertex 
$T^{\mu \nu,\,ab}_{ggH}(k_1,k_2)$ for two off-shell gluons having four-momenta
$k_1$ and $k_2$ and color indexes $a$ and $b$ can be obtained from 
the effective Lagrangian~(1):
\begin{equation}
  T^{\mu \nu,\, ab}_{ggH}(k_1,k_2) = i \delta^{ab} {\alpha_s\over 3\pi} \left(G_F \sqrt 2\right)^{1/2} \left[k_2^\mu k_1^\nu - (k_1 \cdot k_2) g^{\mu \nu}\right].
\end{equation}

\noindent
The two-photon decay process $H\to \gamma \gamma$ in the SM proceeds through $W$ boson and fermion
loops. The corresponding $H\to \gamma \gamma$ coupling is determined 
by the effective Lagrangian\cite{27,28}
\begin{equation}
  {\cal L}_{H\gamma \gamma} = {\alpha\over 8\pi} {\cal A} \left(G_F \sqrt 2\right)^{1/2} F_{\mu \nu} F^{\mu \nu} H,
\end{equation}

\noindent
where $F_{\mu \nu}$ is the photon field strength tensor and 
\begin{equation}
  {\cal A} = {\cal A}_W(\tau_W) + N_c \sum_f Q_f^2 {\cal A}_f(\tau_f).
\end{equation}

\noindent 
Here $N_c$ is the color factor, $Q_f$ is the (fractional) electric charge of the fermion $f$ and 
the scaling variables $\tau_f$ and $\tau_W$ are defined by
\begin{equation}
  \tau_f = {m_H^2\over 4m_f^2}, \quad \tau_W = {m_H^2\over 4m_W^2},
\end{equation}

\noindent
where $m_W$ and $m_f$ are the $W$ boson and fermion masses, respectively.
The amplitudes ${\cal A}_f$ and ${\cal A}_W$ are expressed as\cite{27,28}
\begin{equation}
  \displaystyle {\cal A}_f(\tau) = 2 \left[ \tau + (\tau - 1) f(\tau) \right]/\tau^2,\atop {
  \displaystyle {\cal A}_W(\tau) = - \left[ 2\tau^2 + 3\tau + 3(2\tau -1)f(\tau)\right]/\tau^2,}
\end{equation}

\noindent
where the function $f(\tau)$ is given by
\begin{equation}
f(\tau) = 
  \begin{cases}
    \displaystyle \arcsin^2 \sqrt \tau, \quad \tau \leq 1\\
    \displaystyle -{1\over 4} \left[ \log {1 + \sqrt{1 - 1/\tau} \over 1 - \sqrt{1 - 1/\tau} }  - i\pi\right]^2, \quad \tau > 1.
  \end{cases} 
\end{equation}

\noindent
The effective vertex $T^{\mu \nu}_{H\gamma\gamma}(p_1,p_2)$ for two photons having four-momenta
$p_1$ and $p_2$ obtained from the effective Lagrangian~(3) reads
\begin{equation}
  \displaystyle T^{\mu \nu}_{H\gamma \gamma}(p_1,p_2) = i {\alpha\over 2\pi} {\cal A} \left(G_F \sqrt 2\right)^{1/2} \left[p_2^\mu p_1^\nu - (p_1 \cdot p_2) g^{\mu \nu}\right].
\end{equation}

\noindent
Taking into account the non-zero transverse momenta
of initial gluons 
$k_1^2 = - {\mathbf k}_{1T}^2 \neq 0$ and
$k_2^2 = - {\mathbf k}_{2T}^2 \neq 0$ and 
using the effective vertices~(2) and~(8), one can easily derive the 
off-shell amplitude of gluon-gluon fusion 
subprocess $g^*g^*\to H\to \gamma \gamma$. We have obtained
\begin{equation}
  \bar {|{\cal M}|^2} = {1 \over 1152 \pi^4} \alpha^2 \alpha_s^2 \, G_F^2 \, |{\cal A}|^2 {\hat s^2 (\hat s + {\mathbf p}_T^2)^2 \over (\hat s - m_H^2)^2 + m_H^2 \Gamma_H^2 } \cos^2 \phi,
\end{equation}

\noindent
where $\Gamma_H$ is the Higgs boson full decay width,
$\hat s = (k_1 + k_2)^2$, the transverse momentum of the Higgs particle is
${\mathbf p}_{T} = {\mathbf k}_{1T} + {\mathbf k}_{2T}$ and 
$\phi$ is the azimuthal angle between the transverse momenta
of initial gluons. 
Here we take the propagator of the intermediate Higgs boson in the Breit-Wigner form to avoid any 
artificial singularities in the numerical calculations.
The summation over the polarizations 
of produced on-shell photons is carried with the 
usual formula $\sum \epsilon^\mu \epsilon^{* \nu} = - g^{\mu \nu}$.
In according to the $k_T$-factorization
prescription\cite{18,19}, the summation over the polarizations of 
incoming off-shell gluons is carried with
$\sum \epsilon^\mu \epsilon^{* \nu} = {\mathbf k}_T^\mu {\mathbf k}_T^\nu/{\mathbf k}_T^2$.
In the collinear limit, when $|{\mathbf k}_T| \to 0$, 
this expression converges to the ordinary one 
after averaging on the azimuthal angle. In all other respects the 
evaluation follows the standard QCD Feynman rules. 
We note that the expression~(9) is fully consistent
with one derived in\cite{25} (where the subsequent
Higgs decay $H\to \gamma \gamma$ was not considered).

The cross section of inclusive Higgs boson production 
in the $k_T$-factorization approach is
calculated as a convolution of the off-shell partonic cross section 
and the TMD gluon densities of the proton.
Our master formula reads\footnote{This formula is corrected compared to the one from previous version of paper.}:
\begin{equation}
  \sigma=\int\frac{|\bar {\mathcal M}|^2}{32\pi\, (x_1 x_2 s)^2} f_g(x_1,\mathbf k_{1T}^2,\mu^2) f_{g}(x_2,\mathbf k_{2T}^2,\mu^2) d\mathbf p_{1T}^2 d\mathbf k_{1T}^2 d\mathbf k_{2T}^2 dy_1 dy_2 \frac{d\phi_1}{2\pi}\frac{d\phi_2}{2\pi},
\end{equation}

\noindent 
where $f_g(x,\mathbf k_{T}^2,\mu^2)$ is the TMD gluon density of the proton,
$s$ is the total energy, $y_1$ and $y_2$ are the center-of-mass rapidities of decay photons, 
$\phi_1$ and $\phi_2$ are the azimuthal angles of the initial gluons having the fractions $x_1$ and $x_2$
of the longitudinal momenta of the colliding protons. 
From the conservation law one can 
easily obtain the following relations:
\begin{equation}
  \mathbf k_{1T} + \mathbf k_{2T} = \mathbf p_{1T} + \mathbf p_{2T},
\end{equation}
\begin{equation}
  x_1 \sqrt s = |\mathbf p_{1T}| e^{y_1} + |\mathbf p_{2T}| e^{y_2},
\end{equation}
\begin{equation}
  x_2 \sqrt s = |\mathbf p_{1T}| e^{ - y_1} + |\mathbf p_{2T}| e^{ - y_2},
\end{equation}

\noindent 
where $\mathbf p_{1T}$ and $\mathbf p_{2T}$ are the transverse momenta of final photons.

In the numerical calculations we have tested a few different sets of TMD gluon densities. 
First of them, namely CCFM A0 set, is commonly recognized at present and 
widely used in the phenomenological applications. It has been obtained\cite{35}
from the numerical solution of the CCFM gluon evolution 
equation where all input parameters have been 
fitted to 
describe the proton structure function $F_2(x,Q^2)$. 
Very recently new CCFM-evolved TMD gluon density function
(the so-called JH-2013 set 2) has been presented\cite{37}. The corresponding initial parameters
have been derived from the numerical description of latest precision HERA data
on the proton structure functions $F_2(x,Q^2)$ and $F_2^c(x,Q^2)$. 
Beside these CCFM-evolved TMD gluon densities, we will use the one obtained from the 
Kimber-Martin-Ryskin (KMR) prescription\cite{38}. 
The KMR approach is a formalism to construct
the TMD quark and gluon densities from well-known conventional ones. 
For the input, we have used leading-order Martin-Stirling-Thorn-Watt (MSTW) set\cite{39}.

We now are in a position to present our numerical results and discussion.
After we fixed the TMD gluon densities of the proton, the Higgs boson
cross section~(10) depends on the renormalization and 
factorization scales $\mu_R$ and $\mu_F$. Numerically, we set them to be 
equal to $\mu_R = \mu_F = \xi m_H$. 
To estimate the scale uncertainties of our calculations we vary the parameter
$\xi$ between $1/2$ and $2$ about the default value $\xi = 1$.
We set $m_H = 126.8$~GeV, 
the central value of the measured mass in the diphoton channel\cite{4}.
Following to\cite{40}, we set $\Gamma_H = 4.3$~MeV 
and use the LO formula for the strong coupling constant
$\alpha_s(\mu^2)$ with $n_f= 4$ active quark flavors at
$\Lambda_{\rm QCD} = 200$~MeV, so that $\alpha_s (m_Z^2) = 0.1232$.
Note also that we use the running QED coupling
constant $\alpha(\mu^2)$.
The multidimensional integration in~(10) have been performed 
by the means of Monte Carlo technique, using the routine \textsc{vegas}\cite{41}. 
The corresponding C++ code is available from the authors on request\footnote{lipatov@theory.sinp.msu.ru}.

The ATLAS collaboration has performed
the measurements in the kinematical region defined by $|\eta^\gamma| < 2.37$, 
$105 < M < 160$~GeV and $E_T^\gamma/M > 0.35$~$(0.25)$ for the leading (subleading)
photon, where $M$ is the invariant mass of the
produced photon pair\cite{4}.
The results of our calculations are presented in Figs.~1 --- 3 in comparison 
with the ATLAS data. On left panels, the solid histograms are obtained with the JH'2013 set 2 gluon density
by fixing both the factorization and renormalization scales at the
default value, whereas the upper and lower dashed histograms
correspond to the scale variation as described above. 
The dash-dotted histograms and short dashed curves correspond to the
predictions obtained with the older A0 set and KMR gluon distribution, respectively\footnote{Here the 
$K$-factor from\cite{24} is taken into account for the KMR gluon.}.
Note that to estimate the scale uncertainties 
we have used the set JH'2013 set 2$+$ and JH'2013 set 2$-$ instead
of the default gluon density function JH'2013 set 2. These two sets 
represent a variation of the scale used in $\alpha_s$ in the 
off-shell amplitude. The JH'2013 set 2$+$ stands for a variation of $2\mu_R$, while
set JH'2013 set 2$-$ reflects $\mu_R/2$ (see also\cite{36} for more information).
We find that the ATLAS data are reasonably well described by the
$k_T$-factorization approach supplemented with the latest CCFM-evolved gluon density.
However, the predictions based on the older A0 set and KMR gluon distribution tend to underestimate the data.
The sensitivity of predicted cross sections
to the TMD gluon densities 
(at least, in the overall normalization)
is clearly visible for all considered kinematical variables,
that supports the previous conclusion\cite{25}.

In right panels of Figs.~1 --- 3 we plot the matched NNLO + NNLL pQCD 
predictions\cite{16,17} (or NLO ones for $|\cos \theta^*|$ distribution) 
taken from\cite{4} in comparison with our results and the ATLAS data.
One can see that the measured cross sections are typically higher than
the collinear QCD predictions, 
although no significant 
deviation within the theoretical and experimental uncertainties
is observed, as it was claimed\cite{4}. The $k_T$-factorization predictions
at the default scale (with JH'2013 set 2 gluon density) are rather similar to  
collinear QCD results, providing us a agreement with the ATLAS data at the same level.
We note also that the main part of collinear QCD higher-order corrections
(namely, NLO + NNLO + N$^3$LO + ... contributions which correspond to the $\log 1/x$ enhanced 
terms in perturbative series) is presented in our calculations as a part of the CCFM
evolution of TMD gluon densities. These corrections are known to be large: 
their effect increases the leading-order cross section by about $80 - 100$\%\cite{11,12}.
So that, Figs.~1 --- 3 illustrate the main advantage of $k_T$-factorization approach: 
it is possible to obtain in a straighforward manner the
analytic description which reproduces the main features of rather
cumbersome high-order pQCD calculations.

To conclude, in the present note we apply the 
$k_T$-factorization QCD approach to the analysis of first experimental data on 
the inclusive Higgs production taken by the ATLAS collaboration at the LHC
in the diphoton decay channel.
Using the off-shell amplitude of gluon-gluon fusion subprocess 
$g^* g^* \to H \to \gamma \gamma$ and 
CCFM-evolved gluon density in a proton, we obtained a
reasonably well agreement between our predictions and the ATLAS data. 
We demonstrated that further theoretical and experimental studies of Higgs boson 
production at high energies could 
impose a constraints on the TMD gluon densities of the proton.

{\sl Acknowledgements.}
The authors are grateful to S.~Baranov for very useful discussions and comments.
This research was supported by the FASI of Russian Federation
(grant NS-3042.2014.2), RFBR grant 13-02-01060 and the grant of the 
Ministry of education and sciences
of Russia (agreement 8412).
A.L. and N.Z. are also grateful to DESY Directorate for the
support in the framework of Moscow---DESY project on Monte-Carlo implementation for
HERA---LHC.

\newpage

\begin{figure}
\begin{center}
\epsfig{figure=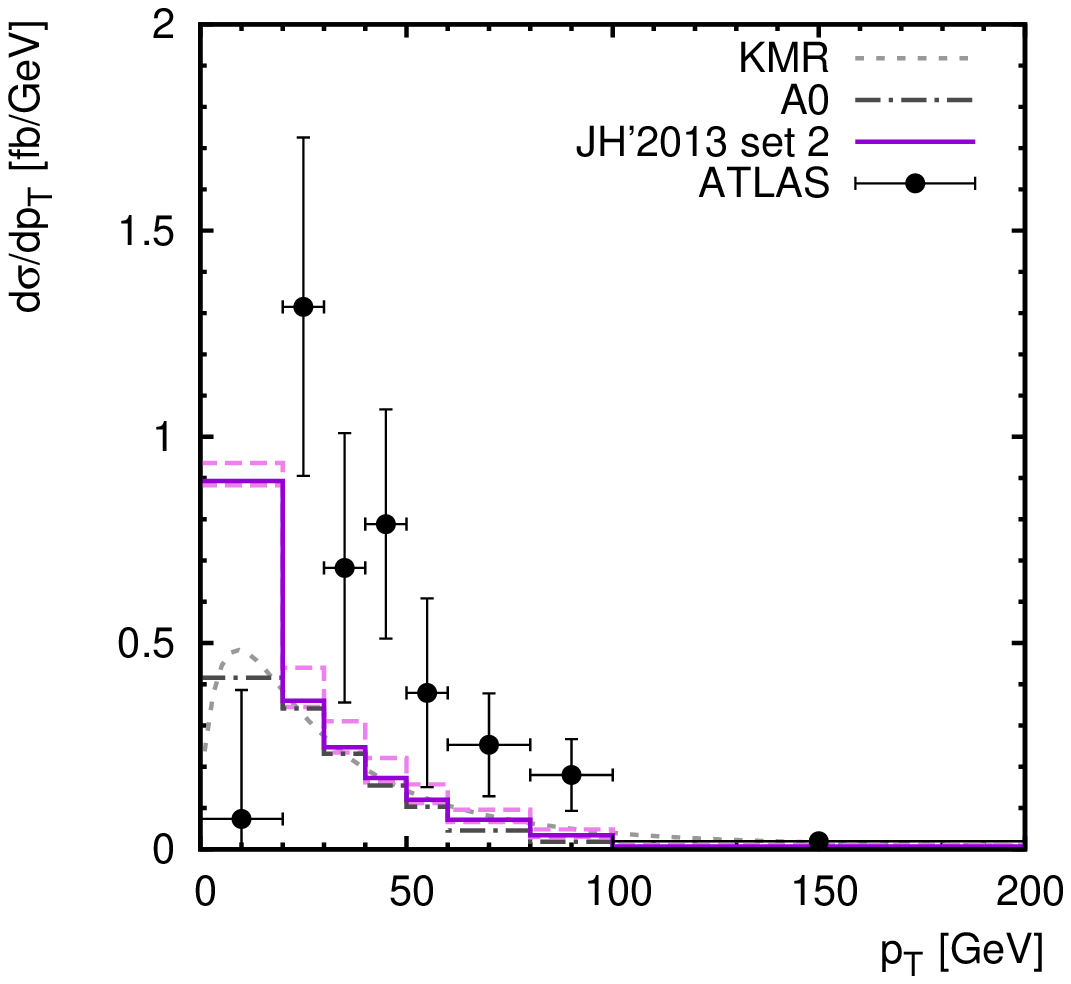, width = 5.5cm, angle = 270}
\epsfig{figure=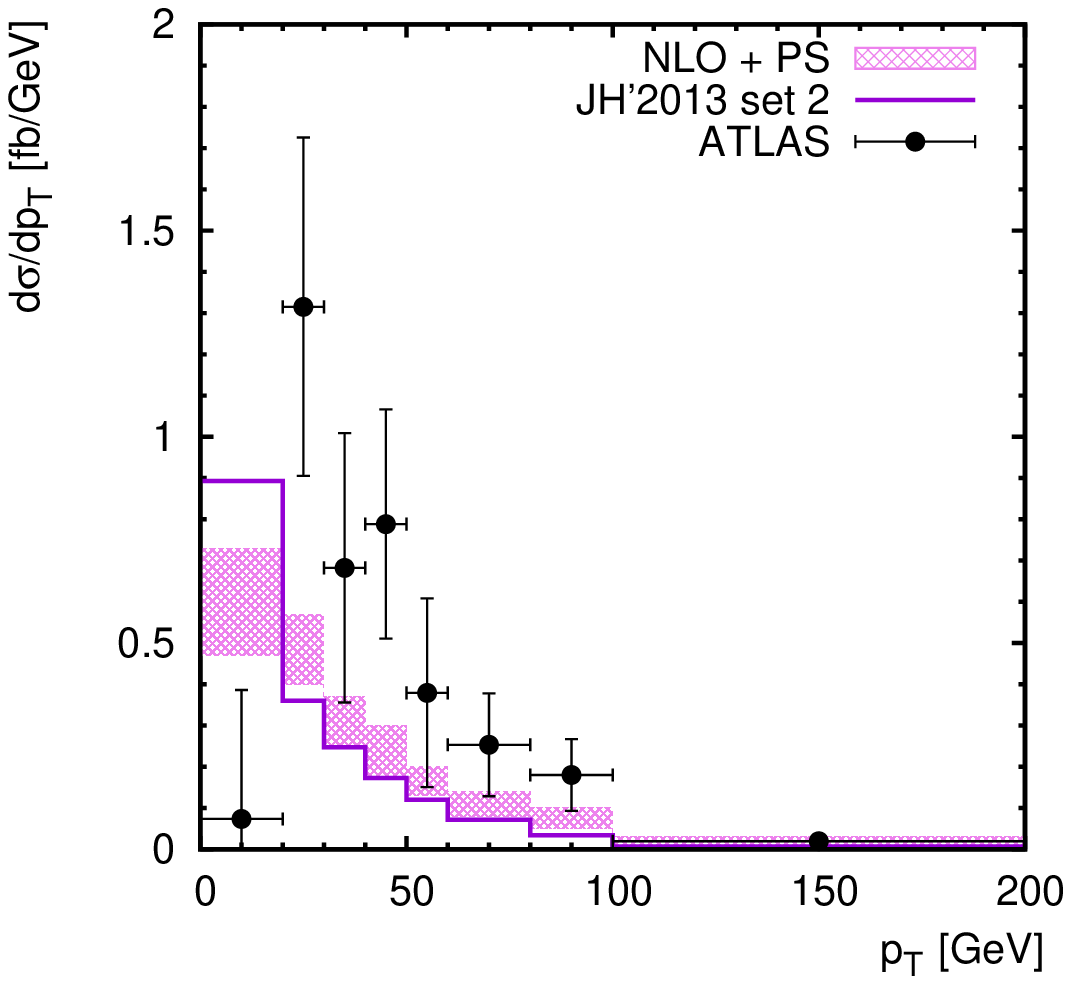, width = 5.5cm, angle = 270}
\caption{The differential cross section of the Higgs boson production in $pp$ collisions 
at the LHC as a function of diphoton transverse momentum. 
Left panel: the solid and dash-dotted histograms correspond to the JH'2013 set 2 and A0 predictions,
respectively; and the upper and lower dashed histograms correspond to the scale variations 
in the JH'2013 set 2 calculations, as it is described in the text. 
The short dash-dotted curve corresponds to the KMR predictions.
Right panel: the solid histogram corresponds to the JH'2013 set 2 predictions,
and the hatched histogram represent the NNLO + NNLL predictions 
obtained in the collinear QCD factorization (taken from\cite{4}). 
The experimental data are from ATLAS\cite{4}.}
\label{fig1}
\end{center}
\end{figure}

\begin{figure}
\begin{center}
\epsfig{figure=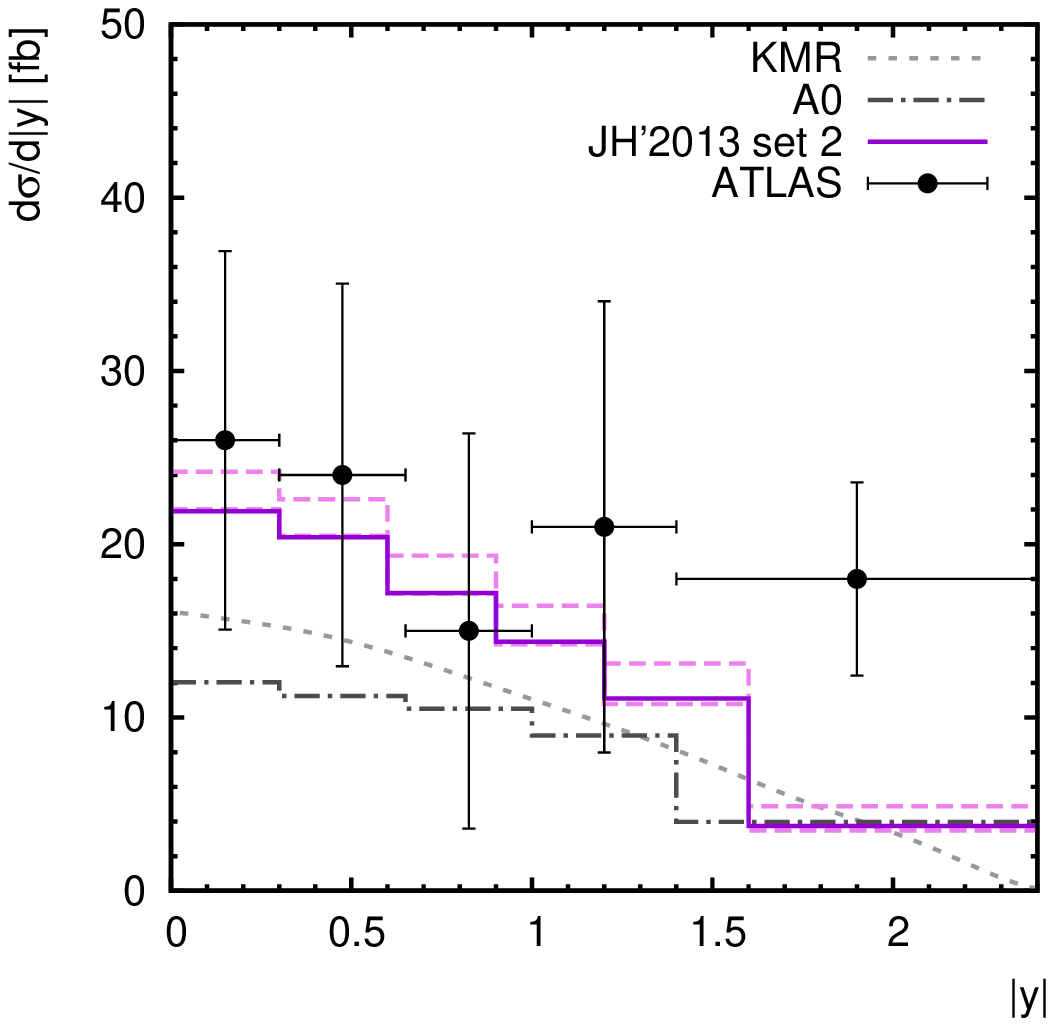, width = 5.5cm, angle = 270}
\epsfig{figure=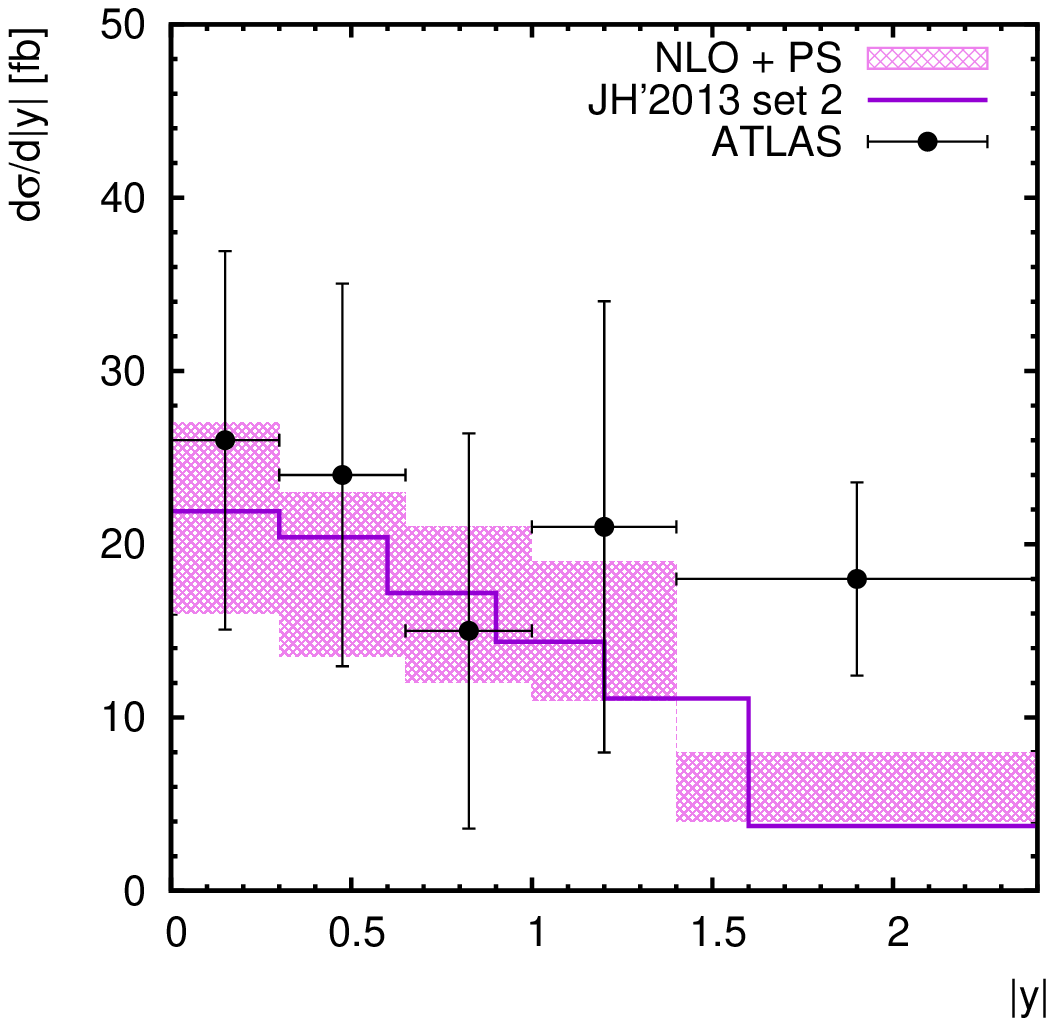, width = 5.5cm, angle = 270}
\caption{The differential cross section of the Higgs boson production in $pp$ collisions 
at the LHC as a function of diphoton rapidity. Notation of all histograms is the same
as in Fig.~1. The NNLO + NNLL pQCD predictions are taken from\cite{4}.
The experimental data are from ATLAS\cite{4}.}
\label{fig2}
\end{center}
\end{figure}

\begin{figure}
\begin{center}
\epsfig{figure=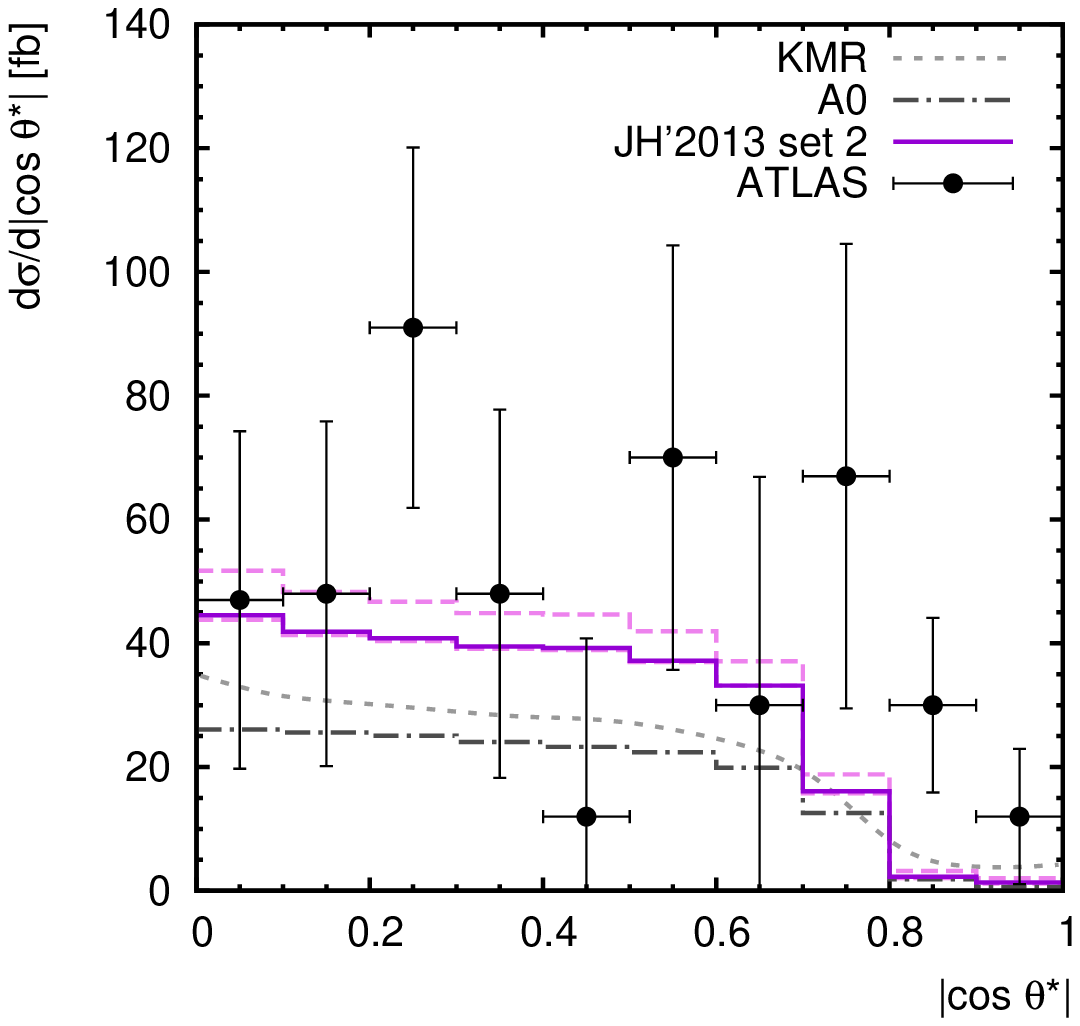, width = 5.5cm, angle = 270}
\epsfig{figure=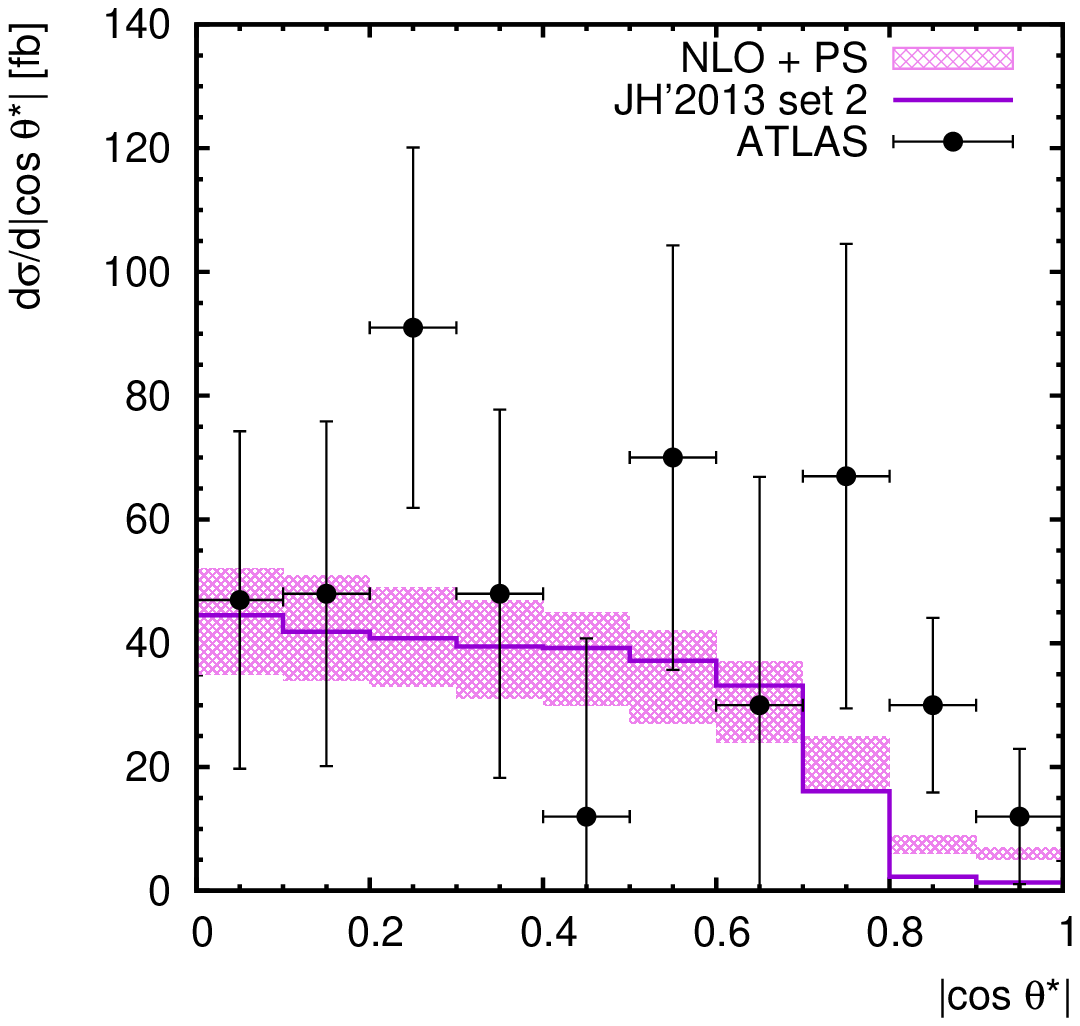, width = 5.5cm, angle = 270}
\caption{The differential cross section of the Higgs boson production in $pp$ collisions 
at the LHC as a function of $|\cos \theta^*|$. Notation of all histograms is the same
as in Fig.~1. The NLO pQCD predictions are taken from\cite{4}.
The experimental data are from ATLAS\cite{4}.}
\label{fig3}
\end{center}
\end{figure}

\end{document}